\documentclass[12pt,a4paper]{article}
\usepackage{graphicx}
\usepackage{rotating}
\usepackage{amsmath,amssymb}
\begin{document}

\begin{center}
\vspace{5mm}
 { \bf
 Comparison of observed and theoretically calculated amplitudes of Forbush 
effects for events associated with the arrival of a shock wave}\\
\vspace{10mm}
 { \bf
 P.G.Kobelev\footnote{Corresponding authors:
e-mail: yanke@izmiran.ru
kobelev@izmiran.ru }, V.N.Zirakashvili, D.N.Dorosheva, V.A.Oleneva, E.G.Klepach, V.G.Yanke$^*$}
\end{center}
% \vspace{-2mm}
{\it \noindent Pushkov Institute of Terrestrial Magnetism, Ionosphere and Radiowave 
Propagation of the Russian Academy of Sciences, 108840 Moscow, Russia}\\

{\small

This paper examines the role of plasma compression regions ahead of interplanetary coronal 
mass ejections (CMEs) in the formation of Forbush effects - decreases in cosmic ray 
intensity during the first phase. The aim of this study is to identify and quantify 
the relationship between the structure of the CME (specifically, the parameters of 
the compression region behind the shock front) and the nature of the observed Forbush effect. 
To this end, events with intensity decrease amplitudes greater than $5\% $ were selected, and a 
theoretical model of cosmic ray behavior at the shock front was tested using modern analytical 
methods and mathematical models. Analysis of the results for two dozen events revealed satisfactory 
agreement between the theory and observations. The key reason for some discrepancies is the need to 
integrate the magnetic field over the scale of the particle gyroradius. The obtained data provide insight into 
the mechanisms of cosmic ray modulation in such environments.}


\section{Introduction}
	
The ability of coronal mass ejections (CMEs) including the shock waves and compression regions 
that precede bodies of CMEs  to dictate the behavior of cosmic rays has been repeatedly demonstrated 
through experimental, theoretical, and statistical research. These disturbances 
trigger a "Forbush effect," which is characterized by a multi-phase structure. The 
sequence of phases includes: precursors, manifesting as a pre-increase and/or 
pre-decrease in cosmic ray (CR) intensity ahead of the shock front; a phase of sharp 
intensity decline, which can sometimes be two-staged; the peak of the effect, 
corresponding to the minimum recorded intensity; and a recovery phase, during 
which CR intensity returns to background levels. A pre-increase is caused by the 
acceleration of particles at the front of an approaching disturbance, resulting 
from their reflection off the shock wave. A pre-decrease occurs when the Earth 
becomes magnetically connected via magnetic field lines to a region of a lower 
particle density; consequently, some cosmic rays from the region behind the 
shock front travel along these field lines and move away from the observation 
zone, which is observed as a pre-decrease in galactic cosmic ray (GCR) intensity prior 
to the main Forbush effect.
	
In cases where the decline profile is two-staged, the first stage corresponds 
to the passage of the compression region (the "sheath") located between the shock wave 
and the CME body, while the second stage corresponds to the passage of the CME body 
itself.

The plasma compression region (sheath) behind the shock front plays a pivotal role, 
as it governs the development of the Forbush effect during its initial phase. It is 
this region  forming behind the shock front  that acts as the "trigger" for changes in 
the structure of the interplanetary magnetic field. This, in turn, leads to the modulation 
of cosmic-ray fluxes  that is, to the occurrence of a Forbush effect.

	The objective of this study is to identify and quantify the relationship between 
the structure of a solar ejection and the formation of an observed Forbush effect during 
its initial stage, as well as to analyze how the plasma compression parameters behind a 
shock front influence the nature of this effect.

\section{Analysis, methodology and event selection}
\subsection{Model for intensity at the shock wave boundary.} 
The first phase of a 
Forbush decrease, associated with the arrival of an interplanetary shock 
wave, typically lasts only a few hours. This duration, in terms of 
its spatial scale, corresponds to the gyroradius of the particles. This 
short time interval can be explained by the reflection of particles from 
a region of enhanced magnetic field located behind the shock front.
	
The jump in the isotropic part of the cosmic-ray (CR) momentum 
distribution function $N(p)$ across the shock front was calculated 
analytically by Webb (1983).
\begin{equation}
N_1-N_2=\frac{2\mu _0^3}{1-\mu _0^2}\frac {j}{v\cos \psi }, 
j=-\frac{V_{sh}p}{3}\frac{\partial N}{\partial p}
\end{equation}
       	
Here, $V_{sh}$ is the shock wave velocity, $v$  is the velocity of the cosmic 
ray (CR) particles, $\psi $ is the angle between the shock front normal and the 
magnetic field upstream of the front, $\mu_0$ is the critical value of the 
cosine of the pitch angle, which corresponds to magnetic reflection and 
is determined by the ratio of the magnetic fields downstream $B_2$ and 
upstream $B_1$ of the shock front, defined as $\mu_ 0^2=1-B_1/B_2$. 
The expression (1) is valid
under the condition that the projection of the cosmic ray anisotropy vector 
onto the shock front normal is zero. This assumption is well justified because  
the observed radial anisotropy of galactic cosmic rays is small. We assume that 
the galactic CR spectrum is a power law function on momentum 
$N(p)=ap^{-(\gamma +2)}$, where $\gamma=2.5$ is the spectral index, and the amplitude $a$ is 
the only free parameter. The differential on energy cosmic ray intensity $J(E)$ is related with the momentum 
distribution  $N(p)$ as $J(E)=p^2N(p)$.
	
Equation (1) demonstrates that the discontinuity in CR density at 
the shock front is directly proportional to the critical pitch-angle 
cosine $\mu _0$ (determined by the magnetic field ratio) and the radial flux $j$ 
(which is related to the momentum derivative of the spectrum).
	In physical terms, this means that particles with pitch angle cosines 
below the critical value $\mu _0$ are reflected as in the magnetic bottle 
downstream of the shock front and cannot penetrate further into the downstream 
region, whereas those with larger pitch angle cosines pass through. The 
difference ($N_1-N_2$) represents a drop in CR number density immediately downstream of the shock front, 
which is observed as the initial phase of the  Forbush decrease.
	
The existence of a jump in cosmic-ray (CR) number density at a 
shock front has been confirmed through numerical solutions of the kinetic 
equation that account for the particle pitch-angle scattering (Ruffolo, 1999; Kirk et al., 1989). 
In this study, the analytical results (Eq. (1)) are compared with observed Forbush decreases.

\subsection{Sample} 
Groups of events were selected from the last three solar activity 
cycles (2000-2006, 2011-2015, and 2023-2026). A total of 20 Forbush effects 
were examined; the selection criteria were limited to a recorded Forbush-effect 
amplitude exceeding $5\% $ and the availability of continuous data on the magnetic 
field and solar wind speed (from the OMNI dataset).  

	No selection was performed based on the magnetic field conditions 
downstream of the shock front, nor was there any filtering based on the angle $\psi $ 
between the shock-front normal and the magnetic field upstream of the front. 
Such events are expected to fall into the category of bad 
cases, as the theoretical framework employed here does not apply under those 
specific parameters. 

\section{Data used}
	The experimental basis for deriving cosmic ray (CR) intensity at an effective 
rigidity of $R_{eff}=10$ GV outside the magnetosphere is the global network of neutron 
monitors, treated as a single, multidirectional detector. Data from this network 
were processed using the Global Spectrographic Method (GSM)
(Krymsky et al. 1966; Nagashima 1971; Belov et al. 1983, 2018), 
which is, in essence, an advanced form of spherical analysis. The method 
accounts for the influence of both the atmosphere 
%(within the "acceptance 
%reception receiving cone" of each detector)
 and the Earth's magnetosphere 
(based on the asymptotic direction of each detector). This yields the 
characteristics of cosmic rays beyond the magnetosphere  specifically, 
hourly estimates of isotropic intensity and of the CR anisotropy vector. 
The data are calibrated against the PAMELA proton flux measurements for 
2009 (Belov et al., 2021) and are available via the (Database GSM, 2026) resource. 
The method ensures data homogeneity and stability, provides good statistical precision 
($\sim 0.01\% $ per hour), and guarantees a completely continuous time series. It serves as a 
powerful tool for solar-terrestrial physics, enabling researchers to use ground-based 
data to look into regions where direct experimental observation is not possible.

\begin{figure}
\begin{center}
\includegraphics[width=15.0cm]{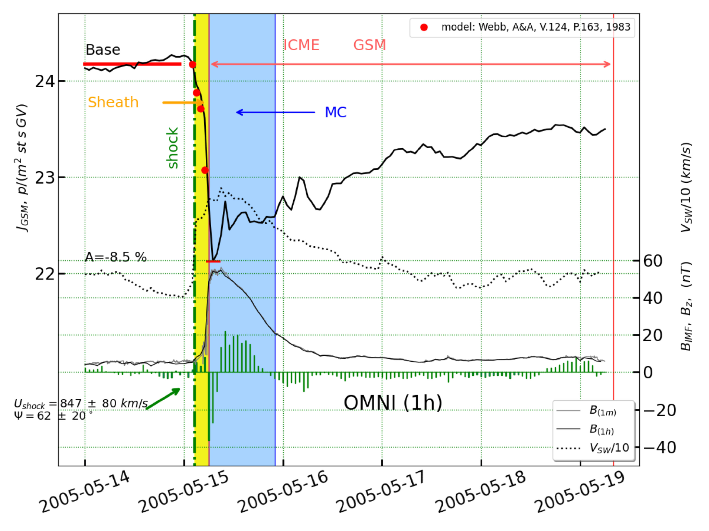}
\caption{Example of good theory-measurement agreement. CR intensity 
according to GSM data (left scale) and interplanetary magnetic field 
induction (right scale) for the Forbush effect 2005-05-16. The right 
scale also shows induction of $z$ - component of magnetic field 
(vertical lines) and solar wind speed (dots). The green vertical 
dashed lines mark the passage of a shock wave associated with an 
interplanetary coronal mass ejection (ICME), while the yellow and 
blue shaded areas correspond to the sheath and magnetic cloud (MC), 
respectively. The area between the two red vertical dashed lines 
corresponds to the ICME. }
\end{center}
\end{figure}

	The following resources concerning interplanetary parameters were utilized in this analysis:
Continuous interplanetary data (magnetic field and velocity) were obtained from the 
OMNI database, which covers solar wind and interplanetary events. These data are 
accessible via the (Database NASA OMNI, 2026) portal. To minimize data gaps, a 
combination of OMNI records with different temporal resolutions  one-minute and 
one-hour  was used. 

\begin{figure}
\begin{center}
\includegraphics[width=15.0cm]{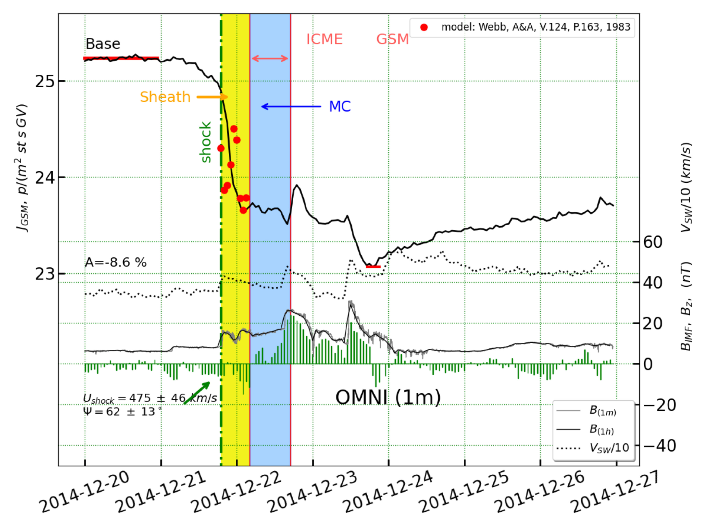}
\caption{Example of satisfactory agreement between theory and measurements for 
the Forbush effect on 2014-12-21 (notation is the same as in Fig. 1).}
\end{center}
\end{figure}

Shock wave velocity, the orientation of the shock surface normal, 
and plasma parameters were extracted from the CfA interplanetary disturbances 
database (Database CfA, 2026). 
Data on heliospheric shock waves and various parameters for each event were 
obtained from the IPShocks database (Database IPShocks, 2026). 
Forbush effects and interplanetary disturbances were sourced from the FEID 
database (Database FEID, 2026). The core concept of this catalog is to 
demonstrate the link between solar wind disturbances and variations of cosmic 
rays for each event. The database covers the period from July 1957 to December 
2026 and includes over 9,400 events, encompassing 120 parameters related to 
cosmic rays, the interplanetary environment, and solar sources.
To determine the angle between the normal to an interplanetary shock wave and 
the Sun-Earth line, data from spacecraft  as presented in established, widely 
recognized catalogs  were used (Database IPShocks, 2026; Database CfA, 2026). 
A description of standard magnetohydrodynamic and kinematic methods, which 
utilize data from multiple spacecraft to calculate the normals to 
interplanetary shock waves, can be found in (Shiryaev et al., 2025). 

\begin{figure}
\begin{center}
\includegraphics[width=15.0cm]{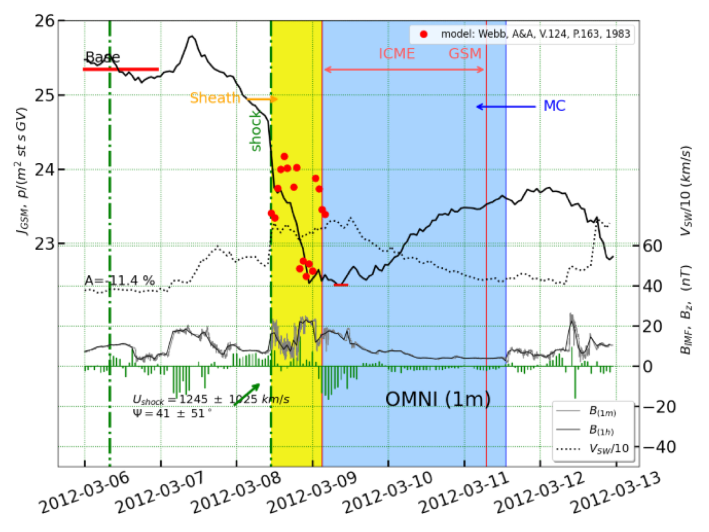}
\caption{Example of poor agreement between theory and measurements for the Forbush 
effect on March 8, 2012 (notation is the same as in Fig. 1).}
\end{center}
\end{figure}

\section{Discussion of results}
	This study examined 20 Forbush decreases with amplitudes 
exceeding $5\% $, occurring during the solar activity maxima of 
cycles 23-25. In summary, the agreement between theory and observation 
was found to be good in 8 cases and satisfactory in 6 cases  the latter 
likely due to the need to integrate the magnetic field over a volume 
defined by the gyroradius. In the remaining cases, specific, anomalous 
features in cosmic-ray behavior were observed during the initial phase 
of the Forbush decrease. Examples of such events are presented below, 
along with a discussion of the potential reasons for any discrepancies. 
	
Fig. 1 illustrates the case of a strong agreement between 
observations and theory (the results are plotted as individual points). 
The duration of the plasma compression region is approximately 4 hours. 
This level of agreement can be attributed to several factors: a large 
angle between the shock front normal and the magnetic field ahead of 
the front, a narrow plasma compression region (lasting only a few hours), 
and a monotonic increase in the magnetic field during this interval.

An example of satisfactory agreement is shown in Fig. 2 - only the 
general course is consistent, but there is a large scatter in the 
theoretical and observed intensity values, which is apparently 
associated with large fluctuations and a non-monotonic course of 
the magnetic field. These reasons can be eliminated by integrating the 
magnetic field over the volume determined by the gyroradius, which 
requires the development of an appropriate technique.

An example of poor agreement is shown in Fig. 3. 
Such agreement should not be expected, as the angle $\psi $  that is 
the angle 
between the shock front normal and the magnetic field ahead of 
the front is small, the theory is not valid for events of this type. 
Furthermore, the event is not entirely isolated: a minor shock wave 
is observed within a 3-day window, and a significant decrease 
(likely a very large diurnal variation) occurs one day prior. 
Additionally, satellite data do not provide a clear determination
 of the angle $\psi $ or the shock velocity $V_{sh}$.

\section{Conclusions}
1) The analysis demonstrates that the analytical expression derived 
by Webb (1983) for the jump of the particle number density at the shock front 
(Eq. (1)) provides a qualitatively accurate description of the 
initial phase of the Forbush effect  specifically, the phase associated 
with the plasma sheath region preceding an interplanetary coronal mass 
ejection (ICME) during the first few hours of the event.

2) Among the events examined, only one-third show good 
agreement between theory and measurements ($\rho > 0.8$). A second 
group  also comprising one-third of the events  exhibits only qualitative 
agreement, characterized by significant scatter. In the third group, 
there is no agreement; this may be due to the event not being isolated, 
the presence of unique features in the Forbush effect, or the fact that 
the magnetic field lines are not tangential to the shock front  meaning 
the angle $\psi $ (between the shock front normal and the magnetic field 
ahead of the front) is small.

3) The approach proposed by Webb (1983) is applicable only during the 
initial period (the first few hours) of the plasma sheath region. To 
provide a complete description of cosmic-ray behavior throughout the 
entire plasma compression region, it is necessary to develop a theory 
applicable to this entire domain. This task is a part of our future plans.

4) The discrepancy between theoretical and observed cosmic-ray intensity 
values for the second group of events could likely be resolved by integrating the 
magnetic field over a volume defined by the gyroradius.  This, however, requires the 
development of an appropriate methodology.

\section*{Acknowledgments}
	The authors are grateful to the participants of the NMDB project 
(http://www.nmdb.eu) and the Russian National Ground-based Network of CR 
Stations (CR Network) (https://ckp-rf.ru/catalog/usu/433536) for the experimental data provided.

\section*{Funding}
	This work was supported by the IZMIRAN budget under 
project 1021025032100209-7-1-1.3.8, "Solar activity and its manifestations 
in the heliosphere and near-Earth space" (FFUM-2026-0005).

\section*{Conflict of interest}
	The authors declare that they have no conflicts of interest with other researchers in this field.

\section*{ORCID iDs}
Kobelev P.G. (https://orcid.org/0000-0002-9727-4395)

Zirakashvili V.N. (https://orcid.org/0009-0001-3524-641X)

%Dorosheva D.N.

%Oleneva V.A.

%Klepach E.G.

Yanke V.G. (https://orcid.org/0000-0001-7098-9094)

\section*{References}
Belov A.V., Dorman L.I., Yanke V.G. (1983). Proc. 18-th ICRC. 10, 144

Belov A.V, Eroshenko E.A, Yanke V.G. et al. (2018) Sol. Phys. 293(4), 68 https://doi.org/10.1007/s11207-018-1277-6

Belov A.V., Gushchina R.T., Shlyk N.S., Yanke V.G. (2021) Bull. Russ. Acad.
 Sci. Phys. 85, 1039 https://doi.org/10.3103/S1062873821090045

Catalog ICME. Catalog near-Earth Interplanetary CME Since 1996. 

Richardson I., Cane H. URL: https://izw1.caltech.edu/ACE/ASC/DATA/level3/icmetable2.htm (accessed: 17.10.2026). 
http://doi.org/10.7910/DVN/C2MHTH 

Database GSM. URL: https://disk.yandex.ru/d/mKHMM2dztqNoHw (part GSM). (accessed: 17.10.2026)

Database NASA OMNI. URL: https://nssdc.gsfc.nasa.gov (accessed: 17.10.2026)

Database of interplanetary shocks. A database of heliospheric shock waves maintained at the 
University of Helsinki. URL: https://ipshocks.helsinki.fi (accessed: 17.10.2026)

Database CfA. Interplanetary disturbances 
database CfA. (Stevens M. L., Kasper J. C.) 
URL: https://lweb.cfa.harvard.edu/shocks (accessed: 17.10.2026) 

Database FEID. Forbush Effects and Interplanetary Disturbances database. 
URL: https://crst.izmiran.ru/w/feid (accessed: 17.10.2026).

Kirk J.G., Heavens A.F. (1989) MNRAS 239, 995 http://doi.org/10.1093/mnras/239.3.995 

Krymsky G.F., Altukhov A.M., Kuzmin A.I et al. (1966) Geomagnetism and Aeronomy. 6(6), 991

Nagashima K. (1971) Rep. of Ionosphere and Space Res. In Japan. 25(3), 189

Ruffolo D. (1999) The Astrophysical Journal. 515, 787 https://doi.org/10.1086/307062 
(https://arxiv.org/abs/astro-ph/9812064v1)

Shiryaev A.O., Kaportseva K.B., Lavrukhin A.S. (2025) A\& AT 35(4), 361

Webb G.M. (1983). Astronomy \& Astrophysics. 124, 163

\end{document}